\documentclass{PoS}

\title{Charmed baryon spectroscopy on the physical point in 2+1 flavor lattice QCD}

\ShortTitle{Charmed baryon spectroscopy on the physical point in 2+1 flavor lattice QCD}

\author{\speaker{Y. Namekawa} for PACS-CS Collaboration,\\
        \llap{} Center for Computational Sciences, University of Tsukuba, Tsukuba,
                Ibaraki 305-8577, Japan\\
        E-mail: \email{namekawa@ccs.tsukuba.ac.jp}}

\abstract{
We investigate the charmed baryon mass spectrum
using the relativistic heavy quark action 
on 2+1 flavor PACS-CS configurations
previously generated on
$32^3 \times 64$ lattice.  
The dynamical up-down and strange quark masses
are set to the physical values 
by using the technique of reweighting
to shift the quark hopping parameters
from the values employed in the configuration generation.
At the physical point, the lattice spacing equals 
$a^{-1}=2.194(10)$~GeV and the spatial extent $L=2.88(1)$ fm. 
Our results for the charmed baryon masses
are consistent with experiments
except for $\Xi_{cc}$,
which has only weak experimental evidence yet.
We also predict mass values for other doubly and triply charmed baryons.
}

\FullConference{The 30 International Symposium on Lattice Field Theory - Lattice 2012,\\
		June 24-29, 2012\\
		Cairns, Australia}

\begin{document}

\section{Introduction}

Recently, a lot of new experimental reports are
delivered on charmed baryons~\cite{PDG_2012}.
BaBar and Belle give very accurate results,
and precision analysis can be accomplished.
In addition, new experiments,
such as J-PARC, PANDA, LHCb, and Belle II
are coming and expected to give further 
insight into charmed baryons.

Mass spectrum of the singly charmed baryons
is determined in the high precision by experiments.
Experimental status of the ground state
is three or four-star,
evaluated by the particle data group.
The excited states are also investigated
fairly well.

In contrast to the singly charmed baryons,
experimental data for the doubly and triply charmed baryons
are not sufficient.
Only one candidate for the doubly charmed baryon, $\Xi_{cc}$,
has been reported by the SELEX Collaboration~\cite{SELEX_2002_2005},
while $\Xi_{cc}$ is not confirmed
in the other experiments such as
BaBar~\cite{BaBar_2006} and Belle~\cite{Belle_2006} groups.
Experimental and theoretical cross checks are needed
to establish $\Xi_{cc}$.
The other doubly charmed baryons and the triply charmed baryon
have not been found by experiment, yet.
Theoretical predictions for the doubly and triply charmed baryons
are helpful to discover these states.

So far, almost all lattice full QCD calculations for charmed baryon spectrum
have been performed
with the staggered dynamical quarks~\cite{Na_2006,Liu_2010,Briceno_2012,Mathur_2012}.
They use some technique,
such as converting the staggered propagators~\cite{Wingate_2003}
or mixed action,
for the valence light quarks
to deal with the tangled flavor structure of the staggered quarks.
Although these approaches have a correct continuum limit,
it breaks the unitarity and complicates the continuum extrapolation.
It is preferred to use the other type of dynamical quarks
which is simple in flavor.
Another point is that their chiral extrapolations
suffer from large higher order corrections.
The pion masses are limited to 220 -- 290 MeV.
NLO SU(2) heavy baryon chiral perturbation theory
is employed to extrapolate their data to the physical point,
but it shows a bad convergence even with $m_{\pi} = 220$ MeV.
It is desirable to perform a simulation directly on the physical point.

The ETMC group studied the charmed baryons
with $N_f=2$ dynamical twisted mass quarks
and Osterwalder-Seiler strange and charm valence quarks~\cite{ETMC_2012}.
They found $\Xi_{cc} = 3.513(23)(14)$ GeV,
which agrees with the SELEX experimental value $\Xi_{cc}^{SELEX} = 3.519(1)$~GeV.
This is the only result that is consistent with the SELEX experiment.
The other lattice QCD simulations show deviations from it.
This disagreement in lattice QCD must be resolved.
A subtle issue in the ETMC calculation
is that the heavy quark mass correction
may not be under control
at their lattice spacings, $a = 0.09-0.06$~fm.
Their results for charmed baryons, especially for the triply charmed baryon $\Omega_{ccc}$,
do not show a clear scaling behavior.
It is better to employ a heavy quark action
that handles mass dependent lattice artifacts in the formulation.
Chiral extrapolation of ETMC data from $m_{\pi}=260$~MeV
using the NLO heavy baryon chiral perturbation theory
is also problematic.

In Ref.~\cite{PACS_CS_2011},
we have shown that the charm quark mass corrections
are under control at $a^{-1}=2.194(10)$~GeV
by adopting the relativistic heavy quark action of Ref.~\cite{RHQ_action_Tsukuba}.
It removes the leading cutoff errors of $O((m_Q a)^n)$
and the next to leading effects of $O((m_Q a)^n (a \Lambda_{QCD}))$
for arbitrary order $n$.
We calculated the spectra of mesons involving charm quarks
using the relativistic heavy quark action
on the 2+1 dynamical flavor PACS-CS configurations
of $32^3 \times 64$ lattice
reweighted to the physical point for 
up, down and strange quark masses.
We found our results are consistent with experiment at a percent level,
and so are those for the decay constants with a few percent accuracy.

Based on this result,
we extend our calculations to the charmed baryon sector.
The notable point is that
our measurements are performed at the physical point.
We are free from the convergence problem
of the heavy baryon chiral perturbation theory.
We check if our masses of singly charmed baryons
reproduce the experimental values
to confirm validity of our calculation.
We also evaluate the doubly and triply charmed baryon spectrum
as our predictions.

%%%%%
\section{Set up}

Our calculation is based on a set of $2+1$ flavor
dynamical lattice QCD configurations
generated by the PACS-CS Collaboration~\cite{PACS_CS_1}
on a $32^3\times 64$ lattice 
using the nonperturbatively $O(a)$-improved Wilson quark action 
with $c_{\rm SW}^{\rm NP}=1.715$~\cite{Csw_NP}
and the Iwasaki gauge action %~\cite{RG}
at $\beta=1.90$.
The aggregate of 2000 MD time units were generated at the hopping parameter
given by $(\kappa_{ud}^0,\kappa_{s}^0)=(0.13778500, 0.13660000)$,
and 80 configurations at every 25 MD time units were used for measurements.
We then reweight those configurations to the physical point
given by $(\kappa_{ud},\kappa_{s})=(0.13779625, 0.13663375)$.
The reweighting shifts the masses of $\pi$ and $K$ mesons 
from $m_\pi=152(6)$~MeV and $m_K=509(2)$~MeV
  to $m_\pi=135(6)$~MeV and $m_K=498(2)$~MeV,
with the cutoff at the physical point estimated
to be $a^{-1}=2.194(10)$~GeV.
Our parameters and statistics
at the physical point are collected in Table~\ref{table:statistics}.

%%%
\begin{table}[t]
\begin{center}
\begin{tabular}{ccccc}
\hline
 $\beta$           &
 $\kappa_{\rm ud}$ & $\kappa_{\rm s}$ &
 \# conf           & MD time
\\ \hline
 1.90              &
 0.13779625        & 0.13663375 &
 80                & 2000
\\ \hline
\end{tabular}
\caption{Simulation parameters.
         MD time is the number of trajectories
         multiplied by the trajectory length.
}
\label{table:statistics}
\end{center}
\end{table}
%%%

The relativistic heavy quark formalism~\cite{RHQ_action_Tsukuba}
is designed to reduce cutoff errors of $O((m_Q a)^n)$
with arbitrary order $n$ to $O(f(m_Q a)(a \Lambda_{QCD})^2)$,
once all of the parameters in the relativistic heavy quark action
are determined nonperturbatively,
where $f(m_Q a)$ is an analytic function
around the massless point $m_Q a = 0$.
The action is given by
\begin{eqnarray}
 S_Q
 &=& \sum_{x,y}\overline{Q}_x D_{x,y} Q_y,\\
 D_{x,y}
 &=& \delta_{xy}
     - \kappa_{Q}
       \sum_i \left[  (r_s - \nu \gamma_i)U_{x,i} \delta_{x+\hat{i},y}
                     +(r_s + \nu \gamma_i)U_{x,i}^{\dag} \delta_{x,y+\hat{i}}
              \right]
     \nonumber \\
 &&  - \kappa_{Q}
              \left[  (1   -     \gamma_4)U_{x,4} \delta_{x+\hat{4},y}
                     +(1   +     \gamma_4)U_{x,4}^{\dag} \delta_{x,y+\hat{4}}
              \right]
     \nonumber \\
 &&  - \kappa_{Q}
              \left[   c_B \sum_{i,j} F_{ij}(x) \sigma_{ij}
                     + c_E \sum_i     F_{i4}(x) \sigma_{i4}
              \right] \delta_{xy}.
\end{eqnarray}
The parameters $r_s, c_B, c_E$ and $\nu$
have been adjusted in Ref.~\cite{PACS_CS_2011}.
It should be noticed that
the parameter $\nu$ is determined non-perturbatively
to reproduce the relativistic dispersion relation for
the spin-averaged $1S$ state of the charmonium.
The heavy quark hopping parameter $\kappa_Q$
is set to reproduce the experimental value of the mass
for the spin-averaged $1S$ state.
Our parameters for the relativistic heavy quark action
are summarized in Table~\ref{table:input_parameters_for_RHQ}.

%%%
\begin{table}[t]
\begin{center}
\begin{tabular}{cccccc}
\hline
 $\kappa_{\rm charm}$  & $\nu$     & $r_s$     & $c_B$     & $c_E$
\\ \hline
 0.10959947            & 1.1450511 & 1.1881607 & 1.9849139 & 1.7819512
\\ \hline
\end{tabular}
\caption{Parameters for the relativistic heavy quark action.
}
\label{table:input_parameters_for_RHQ}
\end{center}
\end{table}
%%%

%-- mass
We use the relativistic operators to obtain charmed baryon spectrum,
because the relativistic heavy quark action is employed in our calculation.
Charmed baryons can be classified under
$4 \times 4 \times 4 = 20 + 20_1 + 20_2 + \bar{4}$.
In addition to $J = 3/2$ decuplet-type 20-plet,
there are $J = 1/2$ octet-type 20-plet and $\bar{4}$-plet.

$J = 1/2$ octet-type baryon operators are given by
\begin{eqnarray}
 O_{\alpha}^{fgh}(x)
 &=& \epsilon^{abc}
     ( (q_f^a(x))^T C \gamma_5 q_g^b(x) ) q_{h \alpha}^c(x),\\
 &&  C = \gamma_4 \gamma_2,
\end{eqnarray}
where $f,g,h$ are quark flavors and $a,b,c$ are quark colors.
$\alpha = 1,2$ labels the $z$-component of the spin.
The $\Sigma$-type and $\Lambda$-type are distinguished as
\begin{eqnarray}
 \Sigma{\rm  -type}  &:& - \frac{ O^{[fh]g} + O^{[gh]f} }
                                { \sqrt{2} },
 \\
 \Lambda{\rm -type}  &:&   \frac{ O^{[fh]g} - O^{[gh]f} - 2 O^{[fg]h} }
                                { \sqrt{6} },
\end{eqnarray}
where $O^{[fg]h} = O^{fgh} - O^{gfh}$.

The decuplet-type $J = 3/2$ baryon operators are
expressed as,
\begin{eqnarray}
 D_{3/2}^{fgh}(x)
 &=& \epsilon^{abc}
     ( (q_f^a(x))^T C \Gamma_+ q_g^b(x) ) q_{h 1}^c(x),\\
 D_{1/2}^{fgh}(x)
 &=& \epsilon^{abc}
     [ ( (q_f^a(x))^T C \Gamma_0 q_g^b(x) ) q_{h 1}^c(x)
     \nonumber \\
 &&   -( (q_f^a(x))^T C \Gamma_+ q_g^b(x) ) q_{h 2}^c(x)
     ] / 3,\\
 D_{-1/2}^{fgh}(x)
 &=& \epsilon^{abc}
     [ ( (q_f^a(x))^T C \Gamma_0 q_g^b(x) ) q_{h 2}^c(x)
     \nonumber \\
 &&   -( (q_f^a(x))^T C \Gamma_- q_g^b(x) ) q_{h 1}^c(x)
     ] / 3,\\
 D_{-3/2}^{fgh}(x)
 &=& \epsilon^{abc}
     ( (q_f^a(x))^T C \Gamma_- q_g^b(x) ) q_{h 2}^c(x),\\
 && \Gamma_{\pm} = (\gamma_1 \mp i \gamma_2)/2, \Gamma_0 = \gamma_3.
\end{eqnarray}

The baryon correlators are calculated
with exponentially smeared sources
and a local sink.
The smearing function is given by $\Psi(r) = A \exp(-B r)$ at $r \not = 0$ and
$\Psi(0)=1$.
We set
$A = 1.2$, $B = 0.07$ for the $ud$ quark,
$A = 1.2$, $B = 0.18$ for the strange quark, and 
$A = 1.2$, $B = 0.55$ for the charm quark.
The number of source points is octupled
and polarization states are averaged
to reduce statistical fluctuations.
Statistical errors are analyzed by the jackknife method
with a bin size of 100 MD time units (4 configurations),
as in the light quark sector~\cite{PACS_CS_1}.
We extract charmed baryon masses by fitting correlators
with exponential functions.

%%%%%
\section{Singly charmed baryon spectrum}

\begin{figure}[t]
\begin{center}
 \includegraphics[width=75mm]{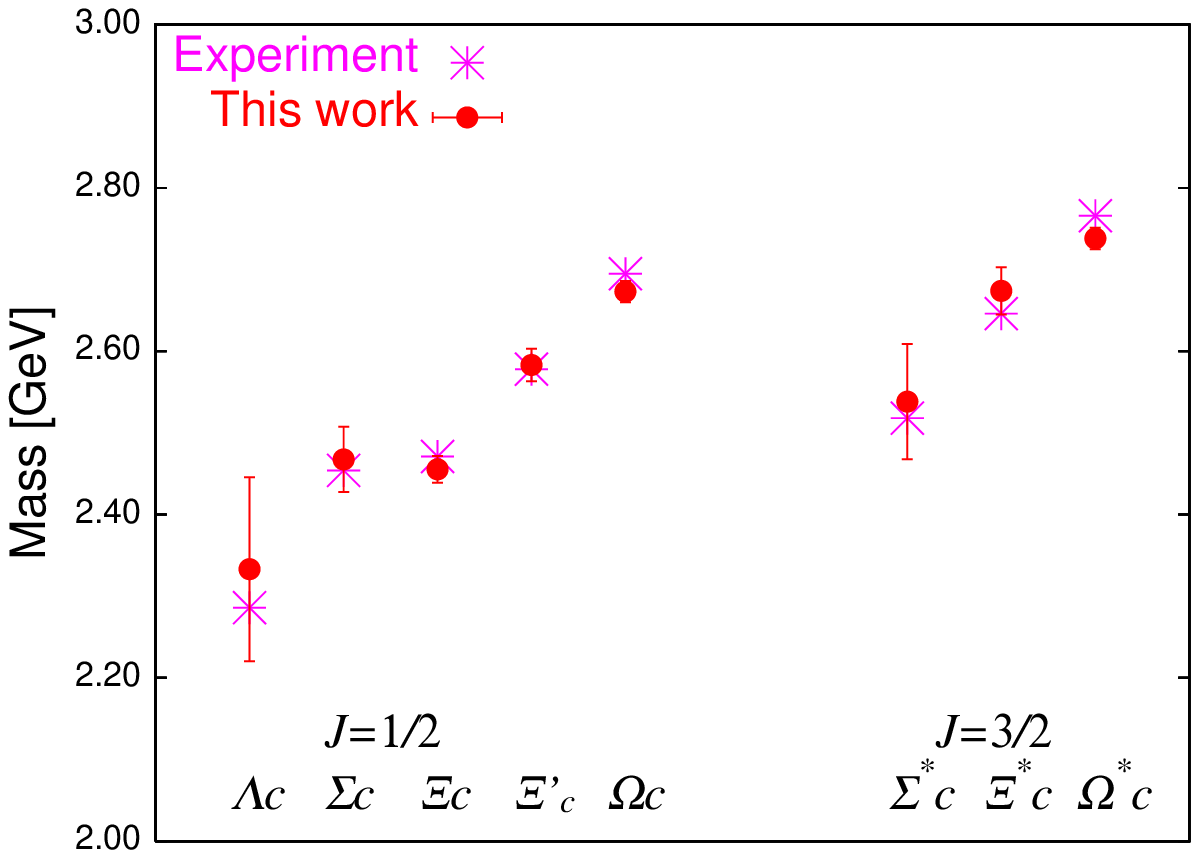}
 \includegraphics[width=75mm]{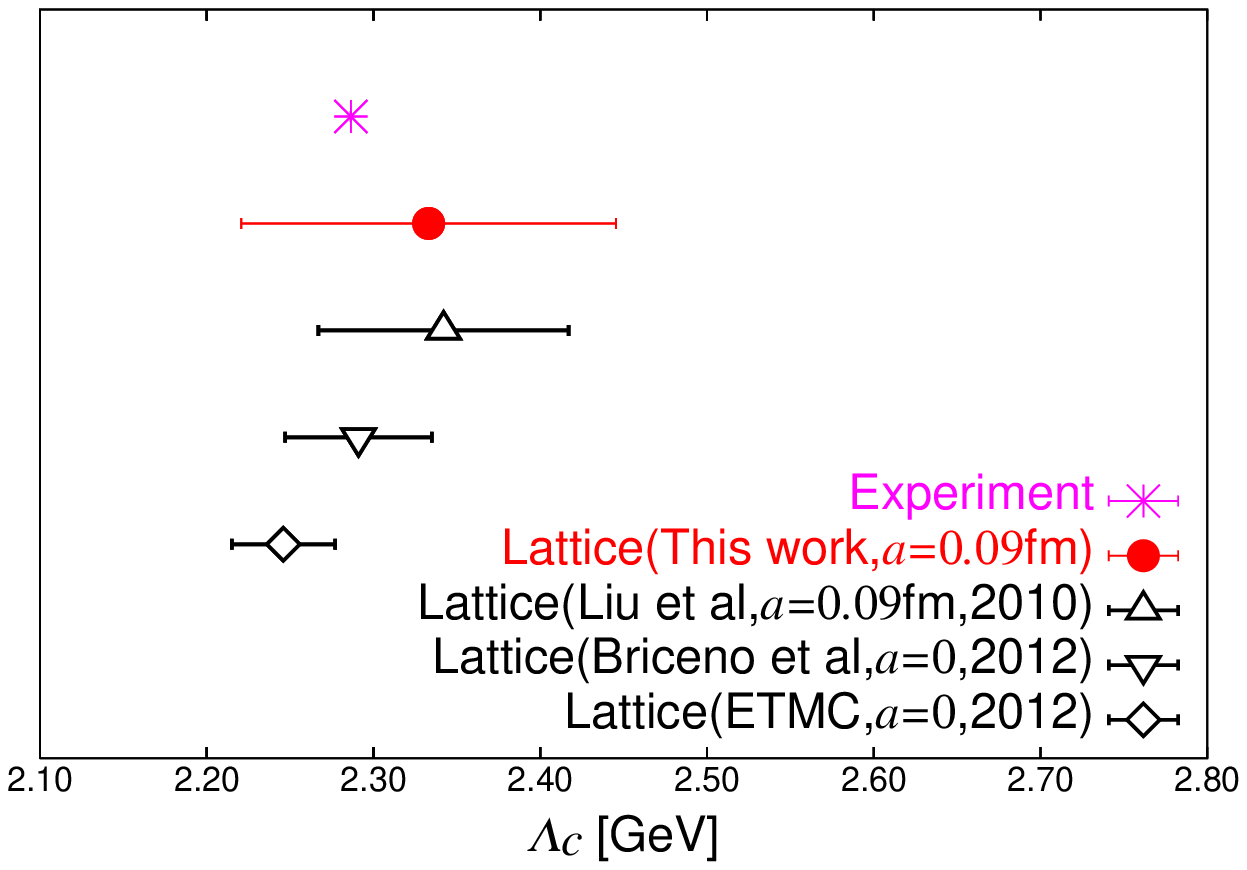}
 \caption{
  Our results for the singly charmed baryon spectrum (left panel),
  and comparison of $\Lambda_c$ mass
  with other lattice QCD results (right panel).
 }
 \label{figure:mass_singly_charmed_experiment}
\end{center}
\end{figure}

Our results for the singly charmed baryon spectrum
at the physical point are summarized in
Fig.~\ref{figure:mass_singly_charmed_experiment}.
All our values for the charmed baryon masses are predictions,
because the physical charm quark mass has already been fixed
with the charmonium spectrum.
We found the predicted spectrum is in reasonable agreement with experiment. 
We also compare our value for $\Lambda_c$
with other results by recent lattice QCD simulations
using the dynamical staggered quarks~\cite{Na_2006,Liu_2010,Briceno_2012},
and the twisted mass quarks~\cite{ETMC_2012}.
All results are consistent with each other.

\begin{figure}[t]
\begin{center}
 \includegraphics[width=75mm]{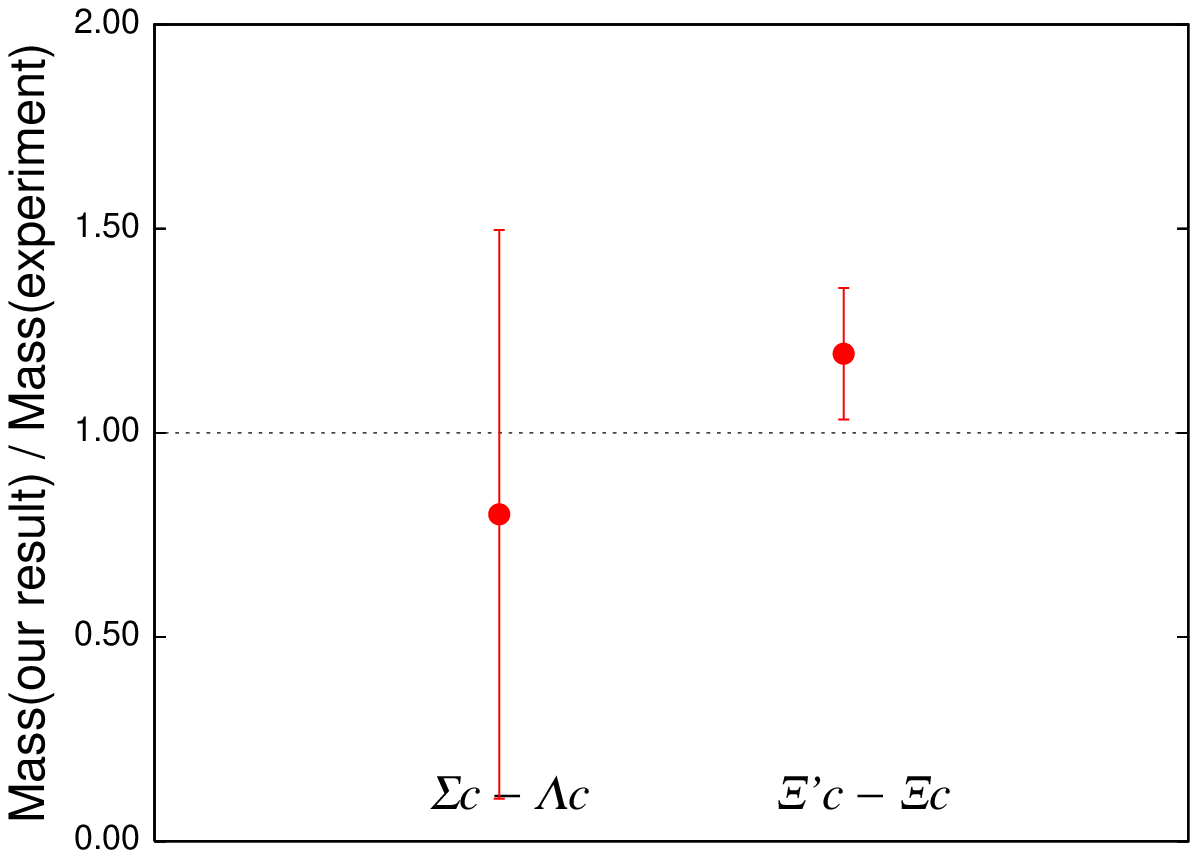}
 \includegraphics[width=75mm]{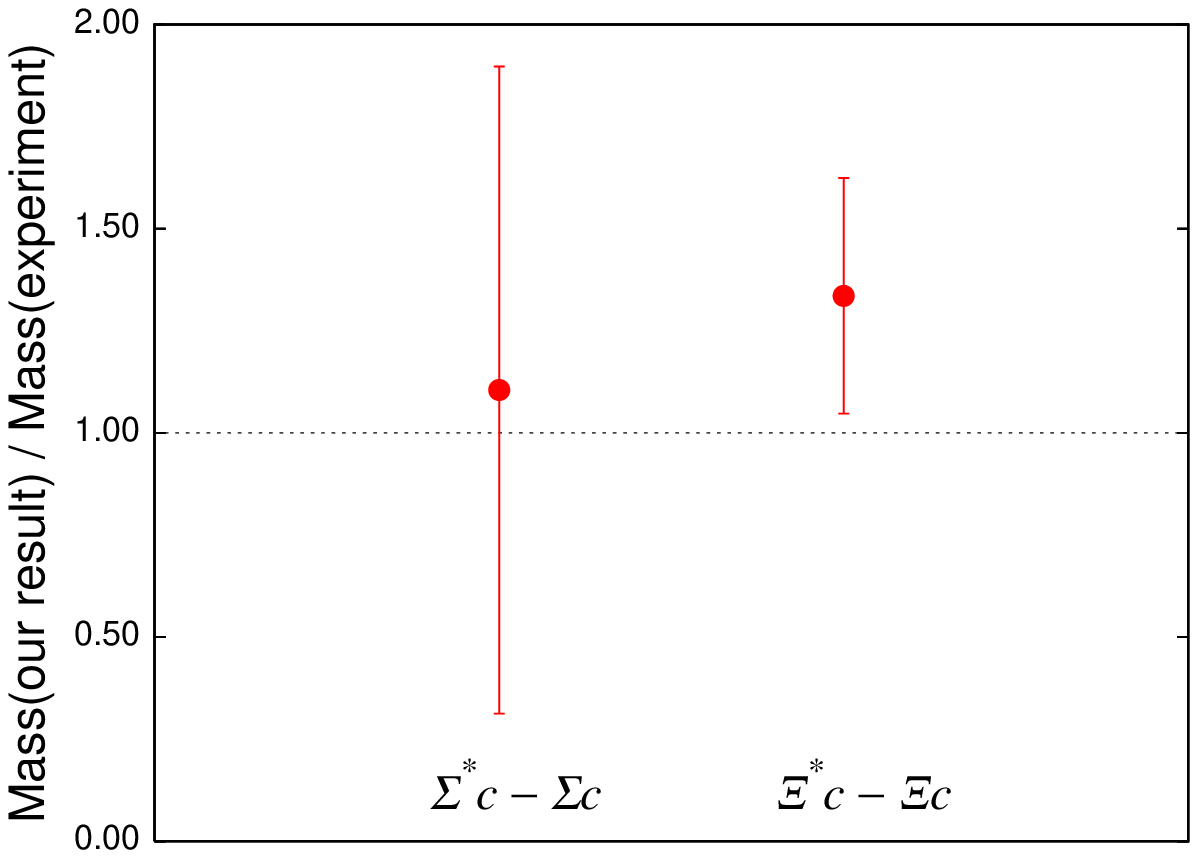}
 \includegraphics[width=75mm]{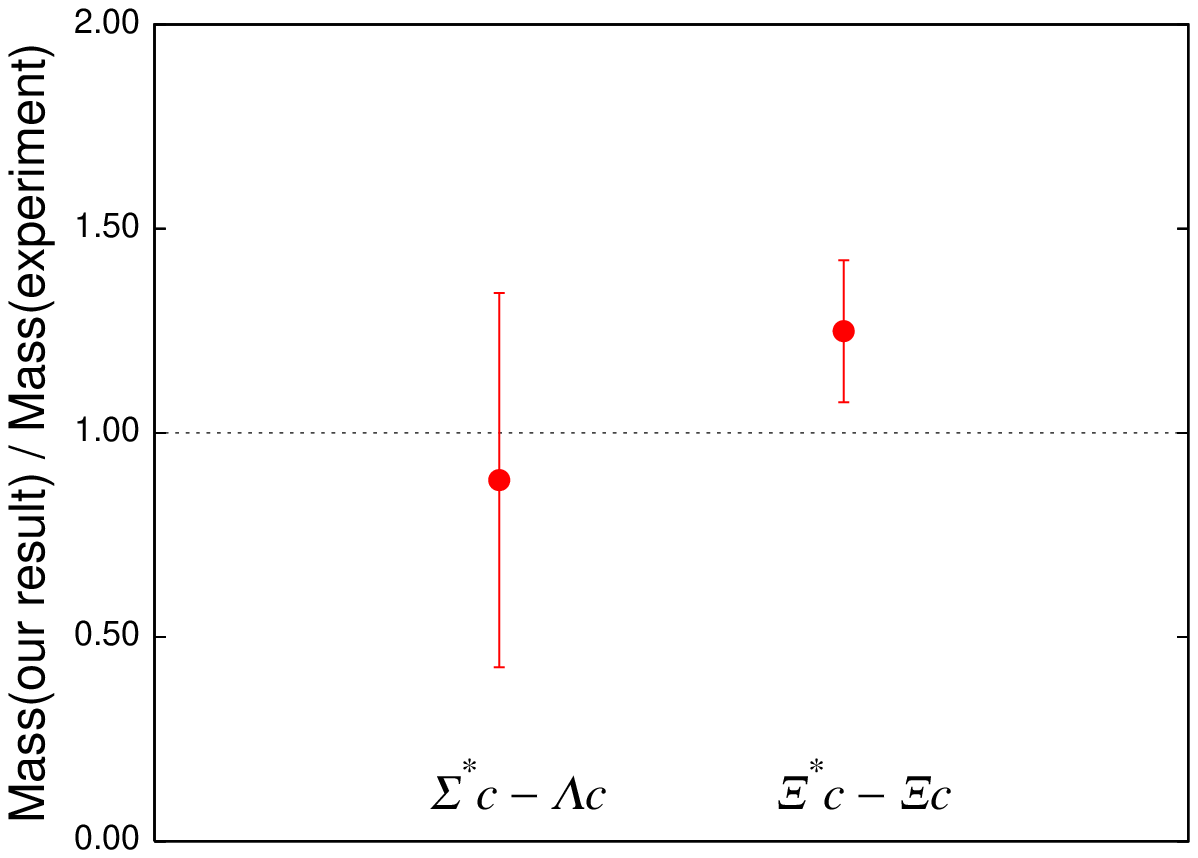}
 \caption{
  Comparison of mass differences of
  $\Sigma_c   - \Lambda_c$ types (upper left panel),
  $\Sigma_c^* -  \Sigma_c$ types (upper right panel),
  $\Sigma_c^* - \Lambda_c$ types (lower panel).
 }
 \label{figure:mass_Sigma_c_Lambda_c_lattice}
\end{center}
\end{figure}

Fig.~\ref{figure:mass_Sigma_c_Lambda_c_lattice}
displays several mass differences.
We have consistent results with experiments in 2 $\sigma$ accuracy.
The decomposition of $J=1/2$ $\Sigma$-type and $\Lambda$-type baryons,
as well as that of $J=3/2$ and $J=1/2$ charmed baryons,
are successful.

%%%

%%%

It is noted that several systematic errors
have not been evaluated, yet.
One is finite size effects.
Though NLO heavy baryon chiral perturbation theory predicts that
finite size effects for charmed baryons are less than 1 \%,
higher order terms can give significant contributions.
A direct lattice QCD check
by comparing spectrum on different lattice volumes is desirable.
Another aspect is that strong decays such as $\Sigma_c \rightarrow \Lambda_c \pi$
are not taken into account.
$\Sigma_c \rightarrow \Lambda_c \pi$ is kinematically prohibited
on our lattice.
Our estimates should be considered as
the lower mass limits for unstable baryons.
Moreover, we have not performed the continuum extrapolation.
A naive order counting implies that the cutoff effects of
$O( \alpha_s^2 f(m_Q a)(a \Lambda_{QCD}), f(m_Q a)(a \Lambda_{QCD})^2 )$
from the relativistic heavy quark action
are at a percent level.
Additional calculations are needed to remove these systematic errors.

%%%%%
\section{Doubly and triply charmed baryon spectrum}

For doubly and triply charmed baryons,
an experimental value has been reported only for $\Xi_{cc}$,
although the experimental status is controversial.
In the other channels,
lattice QCD gives predictions to experiments.

\begin{figure}[t]
\begin{center}
 \includegraphics[width=75mm]{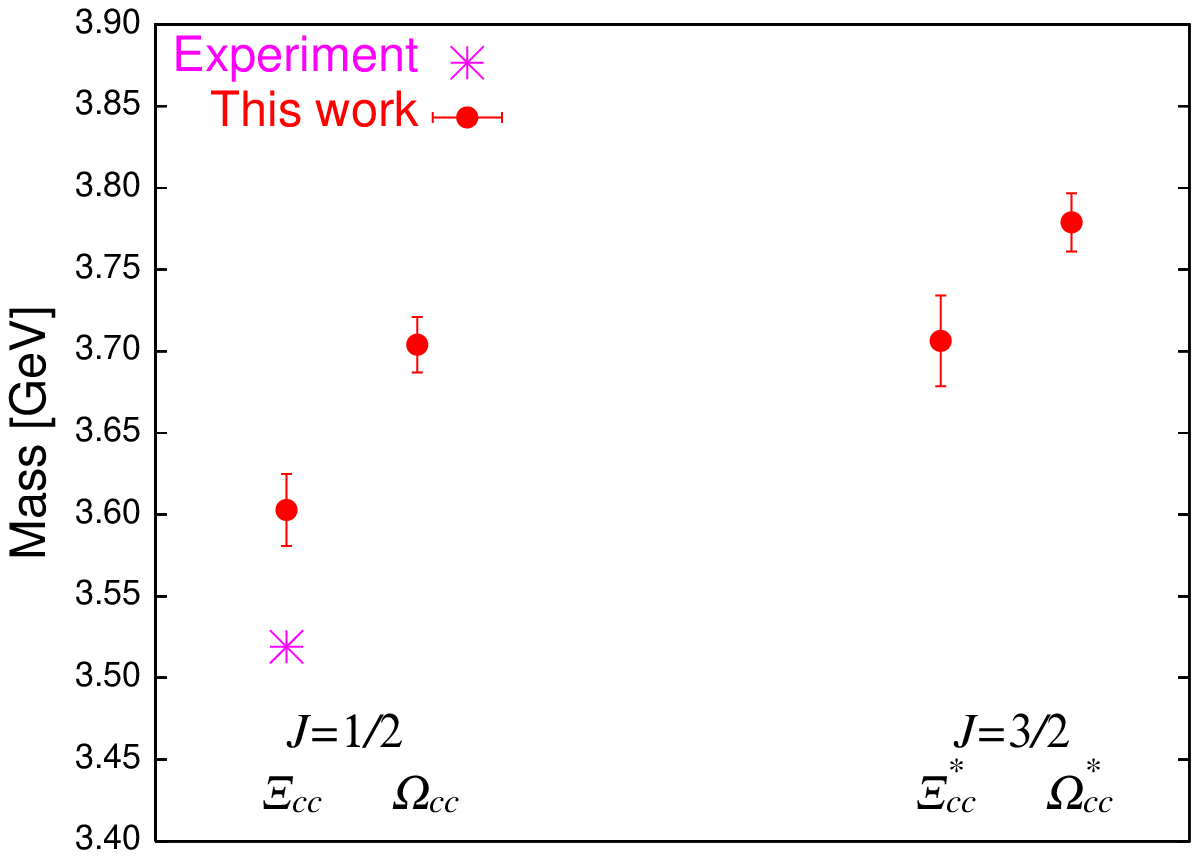}
 \includegraphics[width=75mm]{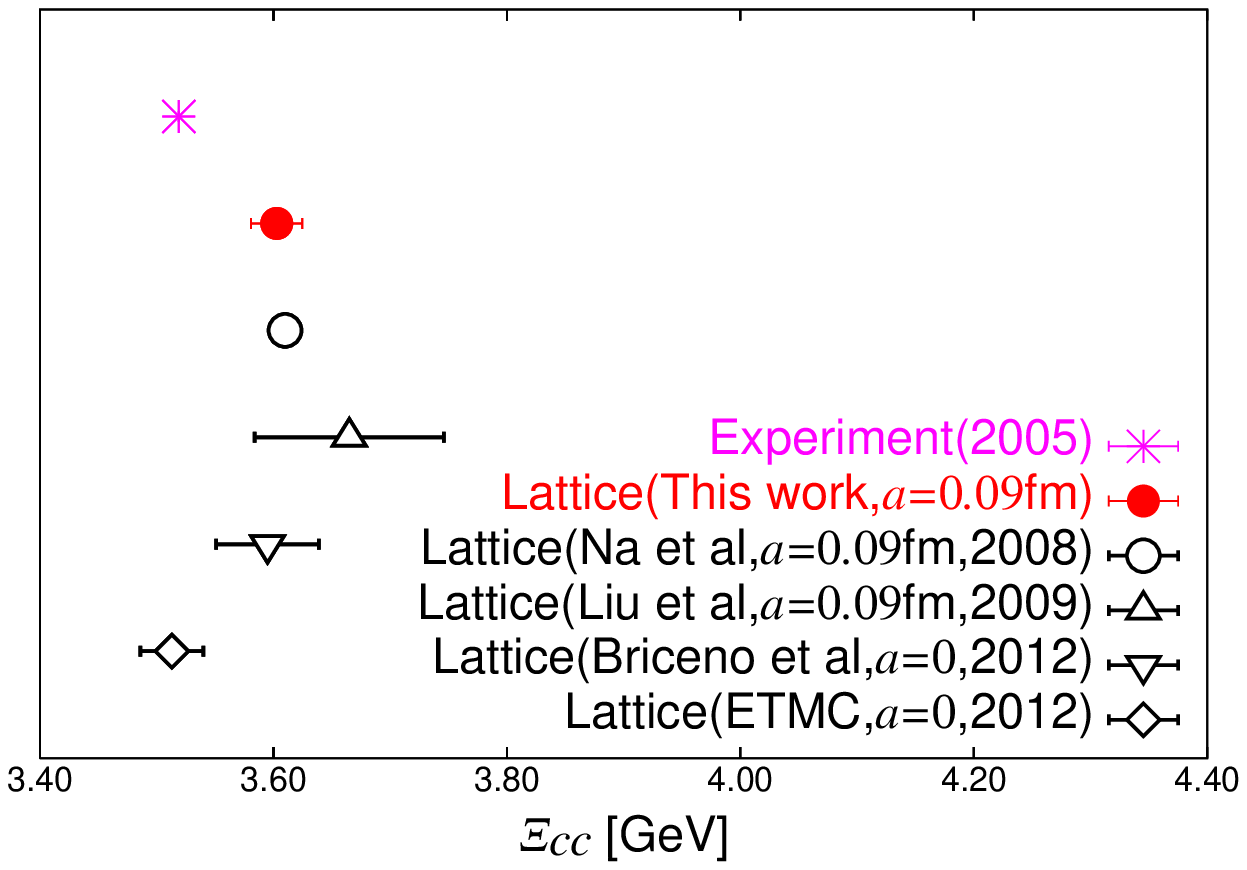}
 \caption{
 Our results for the doubly charmed baryon spectrum (left panel),
 and comparison of $\Xi_{cc}$ with other lattice QCD results (right panel).
 }
 \label{figure:mass_doubly_charmed_experiment}
\end{center}
\end{figure}

Fig.~\ref{figure:mass_doubly_charmed_experiment} shows
our results for the doubly charmed baryons.
Our estimate for $m_{\Xi_{cc}}$ clearly deviates
from the experimental value
by SELEX Collaboration~\cite{SELEX_2002_2005}.
The difference is $4 \sigma$.
We compare our result for $m_{\Xi_{cc}}$
with those by other lattice QCD calculations.
We have a consistent value with other lattice QCD calculations,
except for that by ETMC.

\begin{figure}[t]
\begin{center}
 \includegraphics[width=75mm]{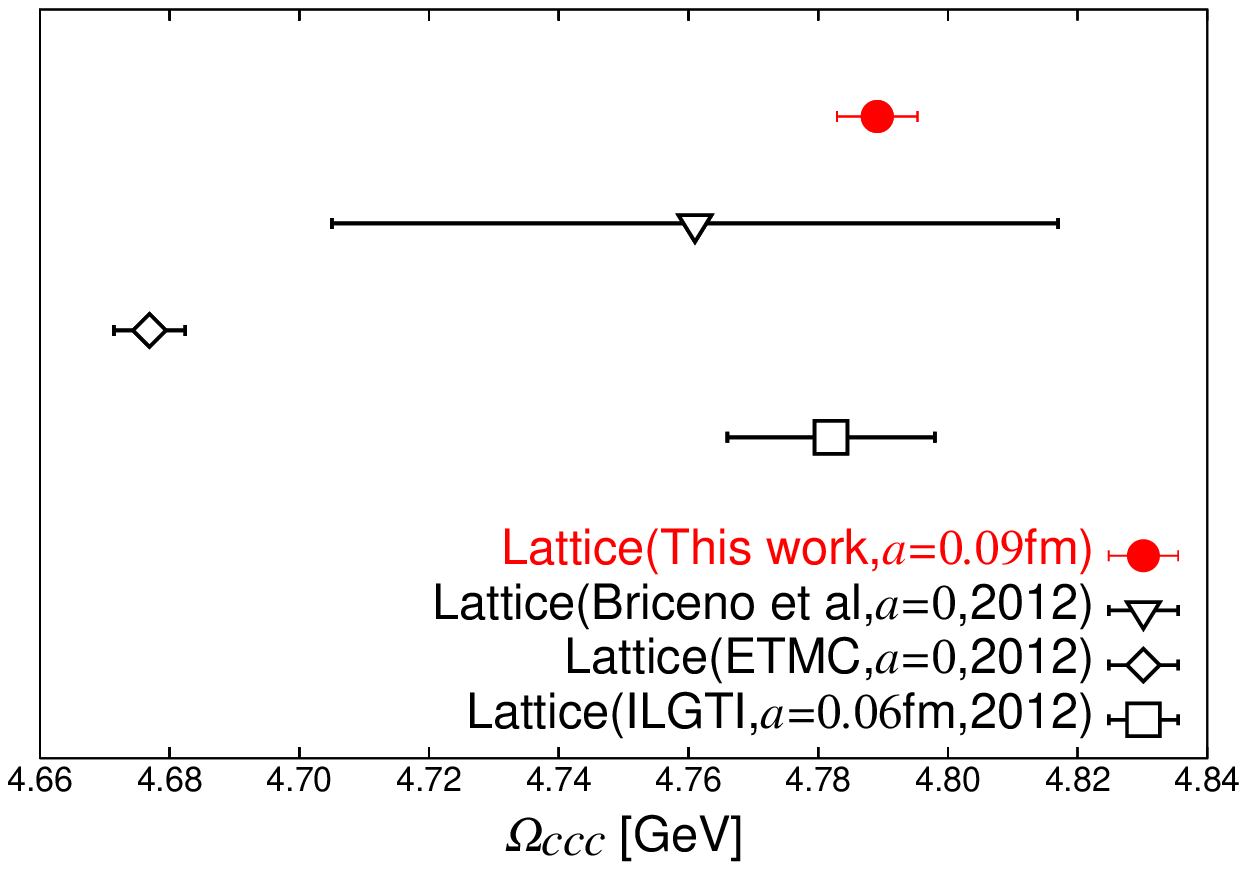}
 \includegraphics[width=75mm]{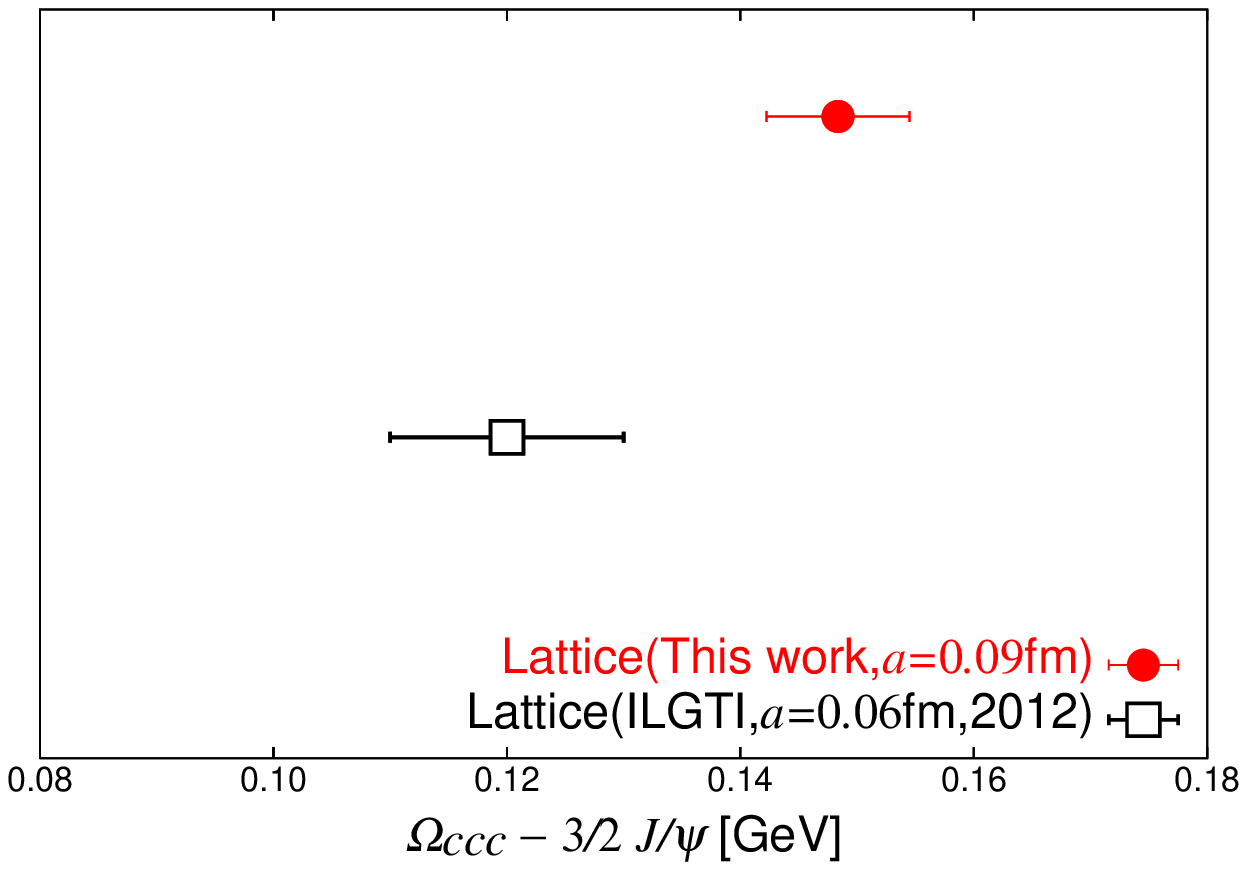}
 \caption{
 Comparison of $\Omega_{ccc}$ (left panel),
 and $\Omega_{ccc}$ $-$ $3/2$ $J / \psi$ (right panel).
 }
 \label{figure:mass_Omega_ccc_lattice}
\end{center}
\end{figure}

Similarly,
Fig.~\ref{figure:mass_Omega_ccc_lattice}
displays lattice QCD results for the triply charmed baryon
from several groups.
Our prediction agrees with those by others,
except for ETMC value.
We also plot $m_{\Omega_{ccc}}$ $-$ $3/2$ $m_{J / \psi}$
mass difference.
A slight discrepancy is observed.
For a more definite comparison,
evaluation of systematic errors
is necessary.
We need to take the continuum extrapolation.

%%%

%%%

%%%%%
\section*{Acknowledgments}

Y.N. thanks Yasumichi Aoki, Carleton DeTar and Nilmani Mathur
for valuable discussions.
Numerical calculations for the present work have been carried out
on the PACS-CS computer
under the ``Interdisciplinary Computational Science Program'' of
Center for Computational Sciences, University of Tsukuba.
This work is supported in part by Grants-in-Aid of the Ministry
of Education, Culture, Sports, Science and Technology-Japan
 (Nos.~18104005, 20105001, 20105002, 20105003, 20105005, 20340047,
 20540248, 21340049, 22244018, 24540250).

%%%%%

\end{document}